%% file: 00-main.tex
\documentclass{sig-alternate}
\usepackage{mathptmx} 

\usepackage{fancyhdr}
\usepackage[normalem]{ulem}
\usepackage[hyphens]{url}
\usepackage[sort,nocompress]{cite}
\usepackage[final]{microtype}
\usepackage[keeplastbox]{flushend}
\usepackage[bookmarks=true,breaklinks=true,letterpaper=true,colorlinks,citecolor=blue,linkcolor=blue,urlcolor=blue]{hyperref}

\usepackage{xspace}
\usepackage{xcolor}
\usepackage{amsmath}
\usepackage{fancybox}
\usepackage{subcaption}

\usepackage{makecell}
\usepackage{ulem}

\usepackage{graphicx}
\usepackage{tikz}

\usepackage{soul}
\usepackage{letltxmacro}
\LetLtxMacro{\oldhl}{\hl}
\renewcommand{\hl}[1]{\oldhl{#1}}               

\input{macros}

\title{Hardware-Software Co-design for \\Broad Acceleration on Commercial PIM Architectures\vspace{-2.5em}} 

\begin{document}

\author{
Johnathan Alsop \\
AMD Research \\
johnathan.alsop@amd.com
\and
Shaizeen Aga \\
AMD Research \\
shaizeen.aga@amd.com
\and
Mohamed Ibrahim \\
AMD Research \\
mohamad1.ibrahim@amd.com
\and
Mahzabeen Islam \\
AMD Research \\
mahzabeen.islam@amd.com
\and
Nuwan Jayasena \\
AMD Research \\
nuwan.jayasena@amd.com
\and
Andrew Mccrabb \\
University of Michigan \\
mccrabb@umich.edu
}





\maketitle
\pagestyle{plain}


\begin{abstract}
\input{01-abstract}

\end{abstract}
\input{02-intro}

\input{03-background}

\input{04-pim-amenability}

\input{05-pim-baseline-eval}

\input{06-pim-optimizations}

\input{07-discussion}

\input{08-related}
\input{09-conclusion}


\bibliographystyle{IEEEtranS}
\bibliography{refs}

\end{document}

%% file: macros.tex
\long\def\ignore#1{}
\sloppypar

\newcommand{\revise}[1]{{#1}} 
\newcommand{\cedit}[1]{{#1}} 
\newcommand{\aedit}[1]{{#1}}

\newcommand{\putsec}[2]{\vspace{-0.0in}\section{#2}\label{sec:#1}\vspace{-0.0in}}
\newcommand{\putssec}[2]{\vspace{-0.0in}\subsection{#2}\label{sec:#1}\vspace{-0.0in}}
\newcommand{\putsssec}[2]{\vspace{-0.0in}\subsubsection{#2}\label{sec:#1}\vspace{-0.0in}}

\newcommand{\figref}[1]{Figure~\ref{#1}}

\newcommand{\secref}[1]{Section~\ref{sec:#1}}

\newcommand*\circled[1]{\tikz[baseline=(char.base)]{
  \node[shape=circle,draw,fill=black,text=white,font=\bf,inner sep=0.5pt] (char)
  {\scriptsize#1};
}}

%% file: 01-abstract.tex
Continual demand for memory bandwidth has made it worthwhile for memory vendors to reassess processing in memory (PIM), which enables higher bandwidth by placing compute units in/near-memory. As such, memory vendors have recently proposed commercially viable PIM designs. However, these proposals are largely driven by the needs of (a narrow set of) machine learning (ML) primitives. 
While such proposals are
reasonable given the growing importance of ML, as memory is a pervasive component, 
there is a case for a more inclusive PIM design that can accelerate primitives across domains.

In this work, we ascertain the capabilities of commercial PIM proposals to accelerate various primitives across domains. We first begin with outlining a set of characteristics, termed PIM-amenability-test, which aid in assessing if a given primitive is likely to be accelerated by PIM. Next, we apply this test to primitives under study to ascertain efficient data-placement and orchestration to map the primitives to underlying PIM architecture. We observe here that, even though primitives under study are largely PIM-amenable, existing commercial PIM proposals do not realize their performance potential for these primitives. To address this, we identify bottlenecks that arise in PIM execution and propose hardware and software optimizations which stand to broaden the acceleration reach of commercial PIM designs (improving average PIM speedups from 1.12x to 2.49x relative \aedit{to} a GPU baseline). Overall, while we believe emerging commercial PIM proposals add a necessary and complementary design point in the application acceleration space, hardware-software co-design is necessary to deliver their benefits broadly.

%% file: 02-intro.tex
\putsec{intro}{Introduction}

As applications of both commercial and scientific importance continue to demand more memory bandwidth, memory vendors are reassessing processing in memory (PIM) as a potential solution. With PIM, in/near memory compute units work in tandem with traditional processors to enable higher effective memory bandwidth (potentially an order of magnitude or more) over that available externally (e.g., to CPUs, GPUs, ASICs, etc.). Recently, multiple memory vendors have proposed commercially viable PIM designs (which we will refer to as "commercial PIM" or simply "PIM") ~\cite{samsungPIM, hynixPIM}.

While proposed commercial PIM designs add a necessary and complementary design point in \cedit{the} application acceleration space, the primary driving factor influencing their design is \aedit{the proliferation of} machine learning (ML) workloads. This is a reasonable strategy given the commercial importance of ML. However, in this work, we \aedit{address} 
a broader question: \textit{how capable are these commercial PIM designs \aedit{of} accelerating important primitives across domains?} We believe this question is worthwhile for several reasons. First, current PIM designs are largely geared toward limited primitives in ML (e.g., dense matrix-vector computations), while ML itself is comprised of a much broader set of (often memory-limited) computations~\cite{metaMLPreprocessing22, googleMLPreprocessing21}. Second, memory is a pervasive component regardless of the processor it is coupled with (e.g., CPU, GPU, ASIC, etc.). As such, even with prioritizing ML, a more inclusive PIM system design is likely to provide both holistic and broad acceleration (ML and non-ML parts of \aedit{an} application). Additionally, a PIM system design with inclusivity in mind is more likely to weather fast-paced workload evolution (e.g., as manifested by ML).

In this work, we ascertain the capabilities of commercial PIM designs to accelerate primitives across domains. We begin with an overview of commercial PIM proposals to establish a strawman PIM system representative of them. Next, we develop a \textbf{PIM-amenability-test} which aids a programmer in assessing if a given primitive is likely to be accelerated by commercial PIM and helps guide the programmer toward an efficient offload of computation to PIM.

Next, we choose to focus on a set of primitives across domains: \textbf{scientific} (wave simulation - \textit{wavesim-volume} and \textit{wavesim-flux} primitives), \textbf{machine learning} (sparse skinny gemms - \textit{ss-gemm}), and \textbf{graph analytics} (\textit{push primitive}). We choose these primitives both for their importance in their source domains and also as, together, they provide a broad set of scenarios for us to assess commercial PIM designs. Further, to assess the benefits PIM provides beyond existing state-of-the-art solutions, we start with GPU-accelerated baselines of these primitives. We then apply our proposed PIM-amenability-test to \cedit{the} primitives under study, which in turn helps us ascertain data-placement and compute orchestration to map the primitives to underlying PIM architecture efficiently. 
We believe that the above process serves as a good template to \aedit{study and map} new primitives to emerging commercial PIM designs.

We then present a performance modelling methodology and observe that, even though \cedit{the} primitives under study are largely PIM-amenable, existing commercial PIM designs do not realize their performance potential for these primitives. This is true even with careful data-placement and compute orchestration. To understand this performance gap, we further analyze the PIM executions and observe that bottlenecks in baseline accelerator executions are exacerbated when \aedit{a} computation is offloaded to PIM (e.g., DRAM row activation overheads). We also observe \aedit{that} PIM acceleration is sensitive to input-dependent cache locality, and how current compute orchestration for commercial PIM designs opens up unique opportunities that \aedit{software can exploit} 
(e.g., sparsity-aware orchestration). 

Following up on our initial vision of a more inclusive PIM design, we identify hardware augmentations and software optimizations to address \cedit{the} above identified challenges and also harness opportunities present. Specifically, we identify three dimensions for our proposed optimizations: \textit{architecture-aware} (mitigating row activation overheads via careful scheduling), \textit{sparsity-aware} (leveraging data sparsity via fine-grain PIM orchestration), and \textit{cache-aware} (allowing input-dependent cache locality via careful PIM offload). Finally, we perform pertinent limit studies anchoring on PIM architecture design decisions and identify their effect on performance. We show how our proposed optimizations stand to broaden the acceleration reach of commercial PIM designs, achieving speedups of up to 2.68x, 3.17x, and 2.43x in scientific, ML, and graph analytics domains respectively (of \aedit{an} available upper-bound of 4x).

Overall, we believe that while emerging commercial PIM designs add a necessary and complementary design point in \cedit{the} application acceleration space, careful attention to \cedit{the} inclusivity of PIM is important. As memory is an omnipresent component in any system, any acceleration that memory can deliver stands to complement 
current (and future) processor-side optimizations. Further, while \aedit{prioritizing ML needs is reasonable in the near term,} 
a more inclusive PIM design stands to holistically accelerate both ML and non-ML computations. Finally, \aedit{hardware-software co-design is necessary to truly and broadly realize PIM acceleration.} 
In summary we believe that, while emerging commercial PIM designs do hold promise, \aedit{we demonstrate that applying PIM to a broader range of primitives can greatly enhance their utility.}

Our work makes the following contributions:
\begin{itemize}
    \item To the best of our knowledge, this is the first work to evaluate emerging commercial PIM designs across primitives from a broad spectrum of domains. 
    \item We develop a \textit{PIM-amenability-test}, comprised of a set of characteristics and associated heuristics, which aids a programmer in assessing if a given primitive is likely to be accelerated by commercial PIM. 
    \item We identify data-mapping and careful compute orchestration for primitives under study and show that, despite such efforts, commercial PIM systems in their current form do not realize their performance potential. 
    \item We identify bottlenecks and opportunities unique to commercial PIM systems and propose hardware augmentations and software optimizations to broaden acceleration reach of commercial PIM designs (improving average PIM speedups from 1.12x to 2.49x relative \aedit{to} a GPU baseline). 
    \item Overall, 
    we make a case for an inclusive PIM design which, while prioritizing commercially dominant primitives (ML), incorporates design changes that enable broad acceleration with PIM. 
\end{itemize}

%% file: 03-background.tex
\putsec{bckg}{Background}
In this section, we first discuss and motivate \aedit{our assumed baseline system}. 
Next, we provide a background on recent commercial PIM prototypes to present a representative strawman PIM design we \aedit{evaluate} in this work. Finally, we conclude with a discussion on the domains and primitives we focus on in this work. 

\putssec{bckg_baseline}{Baseline System - GPU + HBM}

\begin{figure}[t!]
    \centering
    \includegraphics[width=\columnwidth]{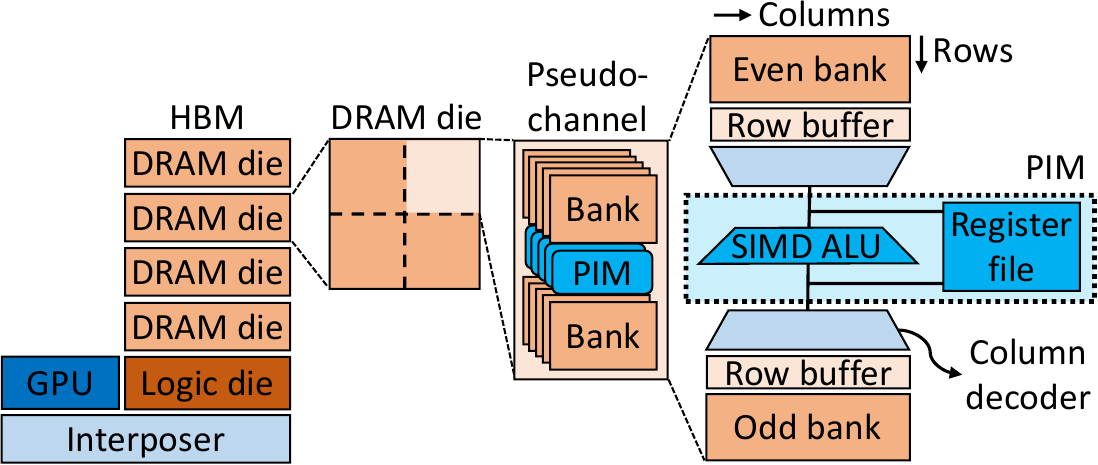}
    \caption{PIM design based on PIM-HBM~\cite{samsungPIM}.}
    \label{fig:pim-background-micro}
    \vspace{-\baselineskip}
\end{figure}

\aedit{The left side of} Figure~\ref{fig:pim-background-micro} 
depicts the baseline system studied in this work: a GPU coupled with HBM memory~\cite{HBM-jedec}. 

\textbf{GPU}: While PIM can be coupled with any processor (CPUs, GPUs), our evaluation assumes a GPU for multiple reasons. First, over the past decade, GPUs have emerged as 
performant and programmable accelerators for a diverse range of highly parallel compute workloads. As such, a GPU processor allows us a strong baseline against which we demonstrate PIM benefits. Second, there exists \aedit{a real} 
PIM prototype~\cite{samsungPIM} coupled with GPUs allowing us a baseline architecture to assess. 
Finally, \cedit{as GPU compute throughput is increasing more rapidly than memory bandwidth, many emerging GPU workloads are likely to be memory bandwidth bound.}

\textbf{High Bandwidth Memory (HBM)}: High bandwidth memory is a specialized form of DRAM that attains high bandwidth and energy efficiency via high density interconnects and 3D stacking. As illustrated in Figure~\ref{fig:pim-background-micro}, each HBM module is a 3D-stack of DRAM dies and a base logic die connected using low-power through silicon vias (TSVs). HBM can be tightly integrated with a processor (in this case a GPU) die on \aedit{a} common substrate such as a silicon interposer~\cite{interposerHBM2016} with an order of magnitude more I/O interconnects~\cite{samsungPIM} than conventional DRAM. 

Each HBM DRAM die is composed of pseudo-channels (pCHs), which further comprise multiple banks that share the data bus associated with a pCH. The address associated with a baseline read/write request specifies the pCH and bank where the data resides along with \aedit{the} row and column address within the bank. On a read request, the specified row is activated in the bank (\textit{row activation}), which causes the data in the row to be read out to the row-buffer associated with the bank (\textit{row open}) after DRAM row activation delay. 
Subsequently, the column decoder selects a DRAM word from the row buffer based on specified column address (\textit{column access}).
Row activation delay overhead can be mitigated in DRAM by exploiting row locality (subsequent accesses to the open row do not incur activation latency) and bank parallelism (column commands to different banks keep the data bus utilized while one bank is activating a row). Note that the basic sequence of operations for \aedit{an} HBM memory access is similar \aedit{to that of} DDR DRAM.

\putssec{bckg_pim}{Commercial PIM Designs}

Recently, two designs for DRAM-attached commercial PIM have been announced by Samsung and SK Hynix. We discuss the details of each design in this section.

\textbf{HBM-PIM}: Samsung's proposed PIM architecture~\cite{samsungPIM,kim2022aquabolt} places ALUs and associated register files on the periphery of DRAM banks (Figure~\ref{fig:pim-background-micro}). 
This design does not disturb the bank and sub-array structures of the memory, improving its viability for commercialization. The PIM ALU is a 256b-wide SIMD datapath that performs sixteen 16b operations in parallel, and is matched to the input/output width of the DRAM cell arrays of the bank. The register files can be used for intermediate results or for staging data from an open DRAM row to reduce the frequency of row activations. Notably, the PIM units do not contain instruction fetch or other "frontend" capabilities, reducing their area costs, and they execute instructions in response to commands issued from the GPU processor. These GPU commands are issued subject to fixed timing constraints, similar to how traditional memory operations are issued. 
The key benefit of this architecture is memory bandwidth amplification, which is achieved by broadcasting the \revise{commands} 
to all banks (or a subset of banks) of a \aedit{pCH 
(normal load/store operations only access a single bank at a time).
This is possible because the data from each bank goes to the associated PIM unit rather than being transmitted across the shared pCH}.
The memory functions as a standard DRAM when the PIM capabilities are not used. 

\begin{figure}[t!]
    \centering
    \includegraphics[width=\columnwidth]{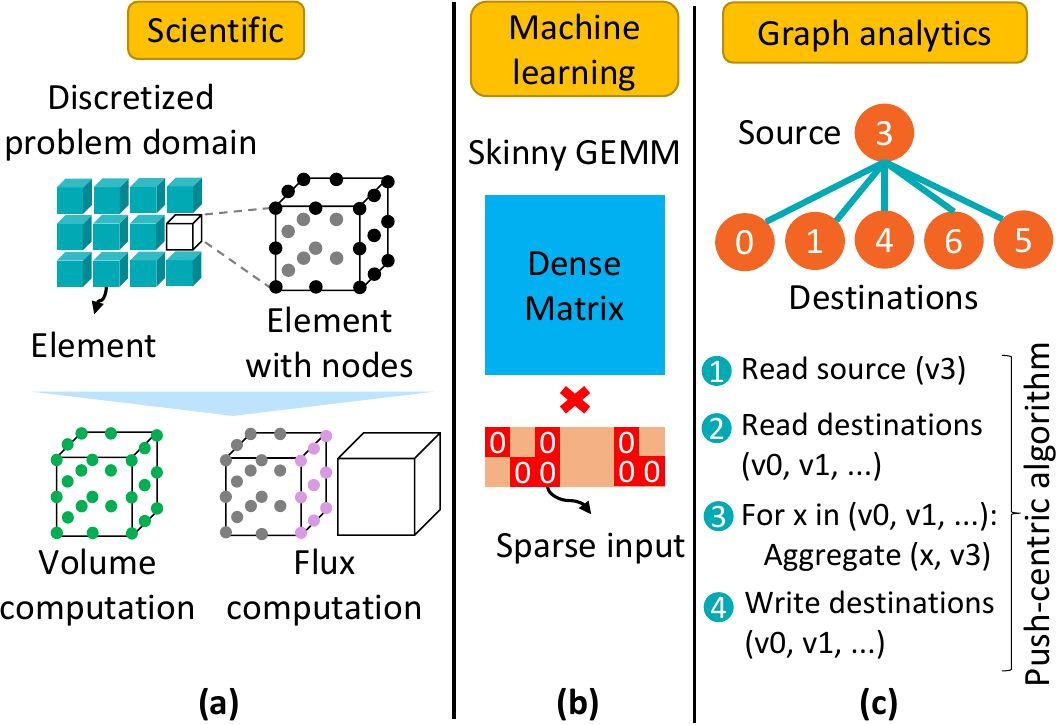}
    \caption{Domains and primitives under study.}
    \label{fig:bg_primitives}
    \vspace{-\baselineskip}
\end{figure}

The authors also describe a prototype implementation named PIM-HBM that is fabricated as an extension to HBM2 and is evaluated in silicon. In the prototype, each PIM unit is shared between a pair of banks, demonstrating the flexibility of the design to provision a PIM unit per bank, per pair of banks, or other grouping of banks to tradeoff performance vs area/cost considerations. The PIM ALUs of the prototype support a limited but generic set of ALU operations, which can presumably be extended in subsequent implementations.

\textbf{GDDR-PIM}: The SK Hynix design~\cite{hynixPIM,lee2019design,he2020newton} describes a PIM system based on GDDR6 that is specifically targeted for ML inference applications and is tailored to matrix-vector multiplications and non-linear activation functions. Despite the specific application focus, this design takes a very similar approach to the \aedit{PIM-HBM} 
\revise{in that it} places compute units on the periphery of DRAM banks and relies on the GPU for instruction triggering in place of native frontend hardware. The datapath is also in a 256b SIMD configuration matched to the bank input/output data width. In the evaluated prototype implementation, a PIM unit is instantiated for each bank of the GDDR6 memory module.

\textbf{Commercial PIM Performance Space}: Table~\ref{tab:commercial_pim_compare} provides key performance metrics for \aedit{the} above commercial PIM designs (per-device compute and data bandwidth). 
We also include state-of-\cedit{the}-art GPU (AMD Instinct\textsuperscript{\texttrademark} MI250 Accelerator) performance metrics (per-HBM stack) for comparison. 
As depicted, PIM data bandwidth is considerably higher than memory bandwidth available to the GPU, while 
GPU compute capability is considerably higher than that of PIM. 

\textbf{PIM Strawman}: For our analyses, we distill a PIM design based on the basic characteristics common to the two memory vendor PIM proposals above but lean closer to HBM-PIM for two reasons. First, HBM-PIM is the more flexible of the two in terms of programmability, and our interest is in further broadening the applicability of PIM. Second, as \aedit{many} modern high-performance GPUs 
use HBM DRAM, HBM-PIM provides a natural comparison point.

\begin{table}[t]
\centering
\caption{Commercial PIMs relative to GPU}
\label{tab:commercial_pim_compare}
\resizebox{\columnwidth}{!}{%
\begin{tabular}{|l|l|l|l|}
\hline
\textbf{Property} & \textbf{MI250-GPU}     & \textbf{HBM-PIM} & \textbf{GDDR-PIM}                                 \\ \hline
Mem clock (GHz) &  1.6  & 1.2 & 1.0                                 \\ \hline
FP16 TFLOPS & 45 & 1.2 & 1 \\ \hline
PIM data BW (GB/s) & n/a & 1229 & 1024 \\ \hline
Mem BW (GB/s) & 400 & 307 & 64 \\ \hline
\end{tabular}%
}
\end{table}

\putssec{bckg_domain_primitives}{Domains and Primitives}
The goal of our work is to assess commercial PIM designs across \aedit{a broader} spectrum of domains than hitherto studied. To this end, we depict the domains and our chosen primitives
in Figure~\ref{fig:bg_primitives}.  We assume 16bit data-type\aedit{s} in our work along the lines of support in HBM-PIM design~\cite{samsungPIM} (See Section~\ref{sec:discussion} for implications to other data types).

\putsssec{bckg_workloads_wavesim}{Scientific - Wave Simulation}
The solution of partial differential equations (PDEs) is critical to many large-scale problems in HPC systems. One such use case, wave simulation, requires solving the wave equation to model the propagation of waves through different media and is used extensively in domains including medical imaging, earthquake modeling, oil and gas exploration, and antenna and radar modeling.

The Discontinuous Galerkin Method (DGM) is a popular algorithm for wave simulation due to its scalability~\cite{dgsim2010}. Like many PDE solvers, DGM discretizes the wave space into a mesh of elements which are distributed among processors in the system (Figure~\ref{fig:bg_primitives}a). It then iteratively executes a volume computation, a flux computation, along with communication and support computations to model wave propagation. The volume computation (termed \textit{wavesim-volume} primitive) performs computations local to each mesh element; the flux computation (termed \textit{wavesim-flux} primitive) propagates conditions at the boundaries \cedit{(faces)} of each mesh element. These two computations dominate execution time for most simulation tasks and as such we focus on these in our work. 

\begin{figure*}[h]
    \centering
    \includegraphics[width=\textwidth
    ]{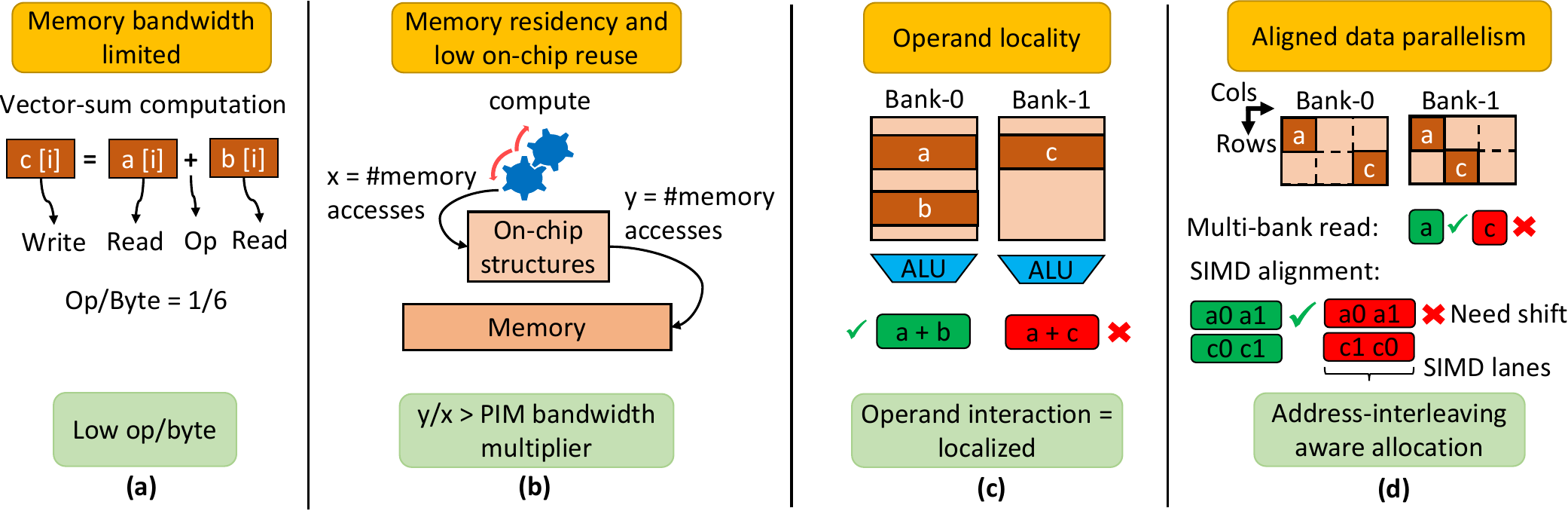}
    \caption{Characteristics of interest for PIM-amenability-test.}
    \label{fig:pim_amenability_test}
    \vspace{-\baselineskip}
\end{figure*}

\putsssec{bckg_workloads_ml}{Machine Learning - Sparse Skinny GEMMs}
Machine learning (ML) continues \aedit{to become ever more pervasive.} 
At the heart of many ML~\cite{PatiIISWC22} workloads is \aedit{General Matrix-Matrix multiplication (GEMM)}. Unlike prior commercial PIM evaluations which focus on skinny GEMMs (the non-shared dimension is small) which are dense, in this work, we focus on GEMMs where one of the matrix inputs is also sparse \aedit{(has many zeros, as in } 
Figure~\ref{fig:bg_primitives}b). We term these sparse skinny gemms (\textit{ss-gemm}) and they manifest in many ML inference scenarios (e.g., Deep Learning Recommendation Model (DLRM)~\cite{naumov2019deep} with small batch sizes). 

\putsssec{bckg_workloads_graph}{Graph Analytics - Push-based Computation}
Graph analytics attempts to derive insights by analyzing the connectivity of graphs and data associated with its edges and vertices. Graph analytics is regularly used for navigation, chemical and biological modeling, social network monitoring and analysis, and many more applications. 

Many common graph analytics workloads operate by iteratively propagating vertex properties (pull or push) across graph edges to neighboring vertices. In pull-based algorithms, a local vertex is processed by reading properties \aedit{from} each of its neighbor vertices and updating the local vertex based on what was read. In push-based algorithms (Figure~\ref{fig:bg_primitives}c), a local vertex is processed by reading its properties and updating neighbor vertices based on what was read \aedit{(using atomic RMWs to avoid race conditions).} 
Push implementations have been found to offer attractive performance properties for many graph algorithms and inputs~\cite{besta2017push,salvador2020specializing}, and they are widely used in GPU graph frameworks~\cite{wang2017gunrock,fu2014mapgraph} and benchmark suites~\cite{nai2015graphbig,davis2015suitesparse}. As such, we focus on \cedit{a} push-based algorithm \aedit{primitive} (termed \textit{push-primitive}).

%% file: 04-pim-amenability.tex
\putsec{pim_amenability}{PIM-amenability}

Given the recency of commercial PIM designs, to effectively and broadly use them, a methodology \revise{is needed} to assess if a given primitive can benefit via PIM acceleration along with guidance on efficiently mapping the primitive to PIM. 
In this section, we present \textit{PIM-amenability-test} which aims to do both, and we describe how we apply it to the primitives under study.

\vspace{-2pt}
\putssec{pim_amenability_test}{PIM-amenability Test}
 We begin this section by discussing a list of characteristics which help evaluate if a given computation is likely to benefit from PIM. We term \aedit{the evaluation of} 
 a given computation against these characteristics as the \textit{PIM-amenability-test} (depicted in Figure~\ref{fig:pim_amenability_test}). We also discuss how this evaluation can aid the programmer in deducing efficient data-placement and compute orchestration \aedit{when mapping} a primitive to PIM. 

Note that our proposed characteristics should be viewed holistically rather than in isolation; that is, manifesting only one of the characteristics does not guarantee PIM acceleration. On the other hand, weakness in a characteristic can in some cases be overcome via optimizations (discussed below). Further, as with any new acceleration solution, while amenability is to be assessed first, efficient computation orchestration is necessary to realize acceleration. We focus in this section on amenability and discuss some of our efficient orchestration learnings in Section~\ref{sec:pim_base_orchestration_general}. 
We only assume in the following discussion that the programmer starts with a state-of-art GPU implementation of the computation. 

\putsssec{pim_amenability_memlimit}{Memory bandwidth limited}
PIM's primary performance benefit comes from increasing effective memory bandwidth.\footnote{Some forms of PIM may offer improved memory latency as well, but this work focuses on the bandwidth benefits.} Therefore, it will not improve performance for workloads that do not stress available memory bandwidth. Memory bandwidth sensitivity will depend on both the workload and the target architecture. This property can be tested analytically by calculating the algorithmic op/byte ratio (Figure~\ref{fig:pim_amenability_test}a) and determining whether it falls in the compute-limited or memory-limited range on a roofline model for the target architecture. 

\vspace{0.5em}
\noindent
\setlength{\fboxrule}{0.2em}
\setlength{\fboxsep}{0.2em} 
\setlength{\shadowsize}{0.1em}
\shadowbox{%
  \begin{minipage}{0.95\linewidth}
    \textbf{PIM-amenable heuristic:} \textit{Low algorithmic op/byte ratio}
  \end{minipage}}

\putsssec{pim_amenability_memlimit}{Memory residency and low on-chip reuse}
Even for workloads that are limited by memory bandwidth, residency and on-chip reuse of computation's inputs/outputs can preclude PIM amenability. Should a computation's inputs/outputs be resident in on-chip structures or manifest enough reuse when moved from memory to on-chip structures, the computation is less likely to attain acceleration with PIM. For the former, employing PIM would necessitate flushing data to memory, resulting in \aedit{added} data movement \aedit{overhead}. For the latter, with enough reuse, moving data closer to the processor to take advantage of the fast and low energy caches and compute available at the processor is a better solution than using slower compute available in memory. 

We can test for both memory residency and low on-chip reuse by determining the proportion of computations that require accessing physical memory \cedit{versus} those that only access on-chip structures (Figure~\ref{fig:pim_amenability_test}b). Comparing this heuristic to \aedit{a} PIM memory bandwidth multiplier (Section~\ref{sec:bckg_pim}) can be used to test if an application manifests this PIM-amenable characteristic. Application dataflow can indicate if a computation's inputs/outputs are likely to be resident in on-chip structures. Computations benefiting from cache-aware optimizations (e.g., tiling, kernel fusion) will also manifest good on-chip reuse. 

\vspace{0.5em}
\noindent
\setlength{\fboxrule}{0.2em}
\setlength{\fboxsep}{0.2em} 
\setlength{\shadowsize}{0.1em}
\shadowbox{%
  \begin{minipage}{0.95\linewidth}
    \textbf{PIM-amenable heuristic:} \textit{Ratio of memory accesses and on-chip structure accesses > PIM bandwidth multiplier}
  \end{minipage}}

\putsssec{pim_amenability_operand_locality}{Operand Locality}
As discussed in Section~\ref{sec:bckg_pim}, compute units in commercial PIM designs are associated with specific memory bank(s). As such, interacting operands in a computation should map to \aedit{the} same bank \aedit{to effectively harness} PIM acceleration (Figure~\ref{fig:pim_amenability_test}c). We term this property \textit{operand locality} in our discussion. In the absence of operand locality, costly GPU-orchestrated data transfers between banks will be necessary\cedit{,} which will eat into PIM acceleration. 

We propose an operand interaction-centric testing of operand locality. Single operand scenarios (e.g., in-place updates) trivially manifest operand locality. Commutative interactions between multiple elements in a single data structure (e.g., reductions) are also trivial, as interactions between elements in the same bank can simply be performed first. 
For multi-operand, multi-data structure cases, we observe that localized operand interactions, wherein small subsets of operands within multiple data structures interact with each other, are generally PIM-amenable. 
A good example of this behavior is element-wise computations (e.g., in vector sum, a single element in each data structure interacts with a single element in another data structure) which can achieve operand locality via careful co-alignment at allocation ~\cite{computeCaches17}. In some cases, localized operand interaction can be induced via data mapping (e.g., packing a matrix in matrix-vector multiplication). In such cases, the costs of doing so have to be factored in when assessing PIM impact. 

\vspace{0.5em}
\noindent
\setlength{\fboxrule}{0.2em}
\setlength{\fboxsep}{0.2em} 
\setlength{\shadowsize}{0.1em}
\shadowbox{%
  \begin{minipage}{0.95\linewidth}
   \textbf{PIM-amenable heuristic:} \textit{Localized operand interaction}
  \end{minipage}}

\putsssec{pim_amenability_regular}{Aligned Data Parallelism}
The bandwidth boost attained in PIM is possible via execution of the same operation in parallel across multiple banks. As discussed in Section~\ref{sec:bckg_pim}, as memory operations have associated row and column addresses, this bank-parallel execution can be employed when operands in different banks in a computation are at the same row/column locations (e.g., see accessing operand \texttt{a} across banks vs accessing operand \texttt{c} in Figure~\ref{fig:pim_amenability_test}d). Note that, within a single DRAM word (256bit single DRAM column), interacting operands therein (e.g., 32bit operands) also have to align (depicted as SIMD alignment in Figure~\ref{fig:pim_amenability_test}d). In absence of this, costly shift operations will be necessary\footnote{Shifter implementations can be costly in DRAM technology due to the limited number of metal layers.}. We term these properties together as \textit{aligned data parallelism}. As processors often spread a contiguous physical address chunk across multiple channels/banks, ensuring interacting PIM operands are interleaved similarly at allocation time can help attain this characteristic.

\vspace{0.5em}
\noindent
\setlength{\fboxrule}{0.2em}
\setlength{\fboxsep}{0.2em} 
\setlength{\shadowsize}{0.1em}
\shadowbox{%
  \begin{minipage}{0.95\linewidth}
 \textbf{PIM-amenable heuristic:} \textit{Address-interleaving aware allocations}
  \end{minipage}}

\putssec{pim_amenability_workloads}{PIM-amenability for Primitives}
We evaluate PIM-amenability of \cedit{the} primitives under study (Section~\ref{sec:bckg_domain_primitives}) using our above test. We also discuss how our test gave good guidance on mapping primitives to PIM. In addition to primitives under study, we also study \cedit{a} \textit{vector-sum} primitive, which has been mapped to commercial PIM 
by prior works. We do so to both evaluate our test against a known PIM-amenable computation and to provide a comparison point for our studied primitives. Finally, our analysis below considers these computations in isolation. That is, dataflow considerations have to be factored-in when other (possibly non-PIM) primitives are involved.

\textbf{Vector Sum:} We depict the \textit{vector-sum} primitive in Figure~\ref{fig:pim_amenability_test}a. This primitive manifests low op/byte (0.17)\aedit{, no data reuse, and} localized operand interaction (\cedit{a} single element in each structure interacts with \cedit{a} single element in other structures).
\aedit{By} co-aligning input structures at memory allocation, \aedit{it} can attain aligned data parallelism. 
As such, 
this primitive is highly PIM-amenable. 

\begin{figure*}[!h]
    \centering
    \includegraphics[width=\textwidth
    ]{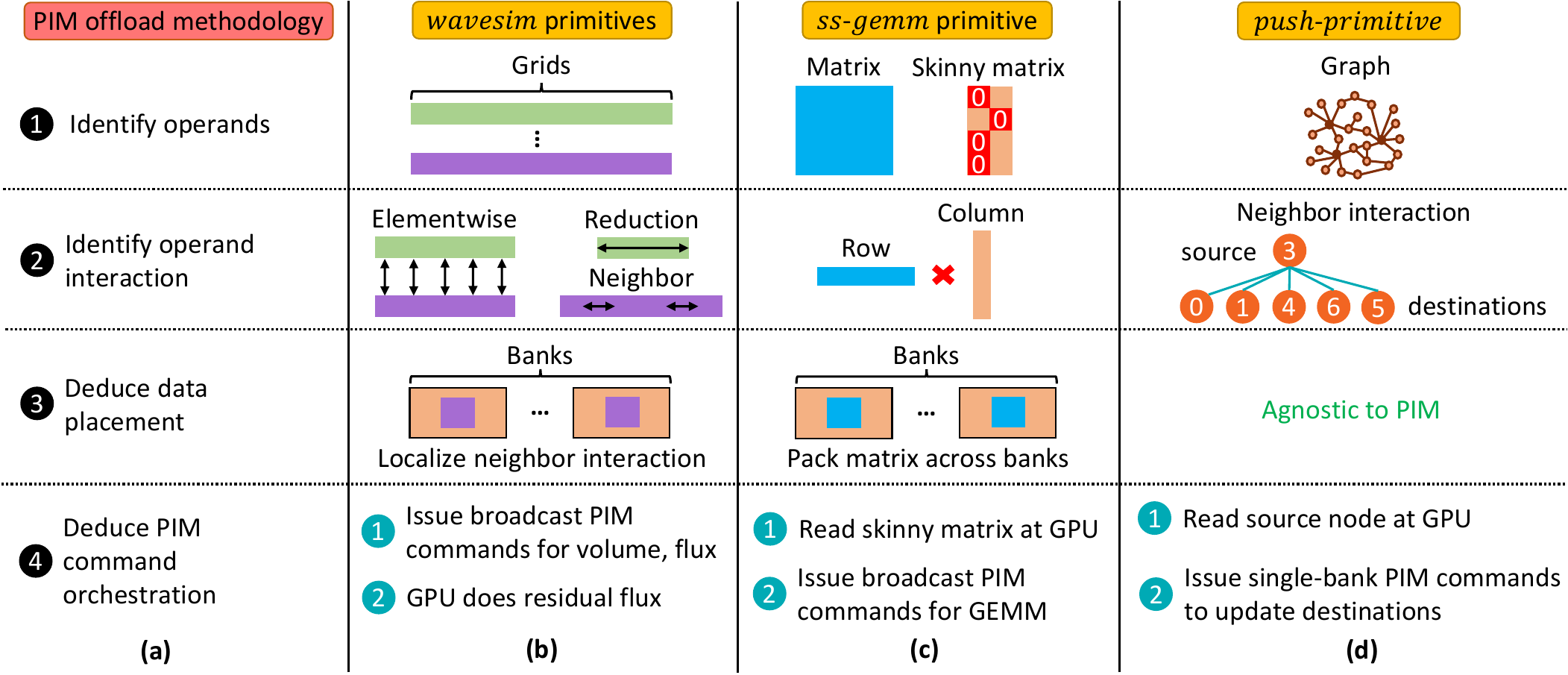}
    \caption{(a) Methodology to offload primitives to PIM. Applying proposed methodology to primitives under study (b, c, d).}
    \label{fig:orch_workloads}
    \vspace{-\baselineskip}
\end{figure*}

\textbf{Wave Simulation:} With low op/byte (0.43-1.72), wave simulation primitives under study (\textit{wavesim-volume} and \textit{wavesim-flux}) are likely to be memory bandwidth-limited for most architectures. Further, for problem sizes that do not fit in cache, these primitives manifest low on-chip reuse. 
While largely relying on localized interaction, these primitives do manifest interactions between neighbor\revise{ing mesh} element \revise{face}s (\revise{flux in} Figure~\ref{fig:bg_primitives}a). As such, 
\aedit{careful memory allocation is necessary to maximize mapping of interacting neighbors to the same DRAM bank}
and further, not all computation can be mapped to PIM (interactions between \aedit{neighboring} elements in different banks). Finally, these primitives operate on large regular grids of elements which can be harnessed to achieve aligned data parallelism. Overall, wave simulation primitives show promise in terms of PIM amenability, albeit care is necessary to attain operand locality.

\textbf{Sparse Skinny GEMMs:} With one of the input matrices being skinny (small $N$ for GEMM $M \times N \times K$), the \textit{ss-gemm} primitive manifests low op/byte (0.5-2 with $N\le4$) and low reuse of data and can benefit from PIM for sufficiently large problem sizes. By streaming the skinny matrix to PIM units and keeping the other matrix stationary in memory, operand locality can be simplified. Further, unless \cedit{the} row-size of \cedit{the} matrix resident in memory is considerably large, aligned data parallelism requires considerable care for \textit{ss-gemm} and we discuss how this guided our data-mapping (\secref{pim-baseline}).

\textbf{Push-based Computation}: The dominant computation for \textit{push-primitive} is the access and update of neighboring nodes (Figure~\ref{fig:bg_primitives}c) which manifests low op/byte (0.25). Sufficiently large problem sizes can stress on-chip structures, albeit on-chip reuse is dependent on connectivity of the graph, leading to selective offloads to PIM. With memory-resident graph (neighbor nodes) and streaming source node data, operand locality is trivial. However, irregularity of accesses to neighbor nodes precludes \aedit{most} aligned data parallelism. As we will discuss, this limits PIM potential for \textit{push-primitive} (\secref{pim-baseline}).

%% file: 05-pim-baseline-eval.tex
\vspace{-2pt}
\putsec{pim-baseline}{Baseline PIM}
We discuss in this section a general methodology, guided by our \textit{PIM-amenability-test}, to offload a given primitive to PIM. Next, we use it to discuss how we 
offloaded
primitives under study to PIM. We then discuss PIM performance and also detail sources of inefficiency. 

\vspace{-2pt}
\putssec{pim_base_orchestration_general}{PIM Orchestration Considerations} We start with a discussion of some general considerations in offloading computations to PIM. In the context of our system, a computation is offloaded to PIM via launching a \textit{pim-kernel}. This is analogous to a GPU kernel, except a \textit{pim-kernel} issues \textit{pim-instructions}. These instructions cause \textit{pim-commands} to be enqueued at the memory controller. Said \textit{pim-commands} trigger either computations on operands in DRAM, data movement between DRAM and registers, etc. In order to preserve register dependencies, \textit{pim-commands} are issued in FIFO order from the memory controller queue. Note that, while we focus on \aedit{the} orchestration of standalone primitives, additional considerations will be necessary (e.g., managing caches) when other computations (non-PIM) are involved that access the same data as PIM primitives (see Section~\ref{sec:discussion}).

\putssec{pim_base_offload}{Offloading Primitives to PIM}
We first discuss our general PIM offload methodology followed by applying it to primitives under study. We also include \textit{vector-sum} discussion for comparative purposes.

\putsssec{pim_base_offload_algo}{PIM Offload Methodology}
While assessing \aedit{the} offloading of a primitive to PIM, as discussed in Section~\ref{sec:pim_amenability}, a programmer can first assesses the primitive using our proposed \textit{PIM-amenability-test}. Assuming PIM amenability exists, we discuss here a methodology which can serve as a general template to offload a primitive to PIM. We depict it in Figure~\ref{fig:orch_workloads}a.

As discussed in Section~\ref{sec:pim_amenability_test}, ensuring operand locality and aligned data parallelism is critical to attaining PIM acceleration. To that end, identifying operands or data structures (\circled{1}) and, more crucially, identifying interactions between them (\circled{2}) are the first steps in offloading to PIM. Subsequently, operands are placed in DRAM banks (\circled{3}) such that costly inter-bank communication, cross SIMD operations, and (where possible) inter-row interactions are avoided. Finally, (\circled{4}), a stream of \textit{pim-instructions} is deduced to orchestrate the computation over PIM units. 

In many cases, compute orchestration follows naturally once data-placement is deduced. However, we also observe that allocation/management of \textit{pim-registers} to minimize row activation overheads \aedit{(i.e., staging data from open rows into \textit{pim-registers})} is often necessary. 
Note that GPU implementations similarly  optimize for effective use of registers.

\putsssec{pim_base_offload_vc}{Vector Sum}
\textbf{Data placement:} As \textit{vector-sum} manifests elementwise interaction between two input and one output array, 
\aedit{allocating structures such that elements at a given offset are mapped to the same bank~\cite{aga2019co} ensures PIM-amenable data placement.}
Note that this is a common operand interaction scenario for PIM-amenable computations. 

\textbf{Command orchestration:} Command orchestration for \textit{vector-sum} is done via broadcast \textit{pim-commands}\cedit{,} which sequentially read input \cedit{values}, perform add\cedit{s}, and write output \cedit{values}. Further, effective use of \textit{pim-registers} to stage data from DRAM to minimize row activation overheads is needed. 

\putsssec{pim_base_offload_wavesim}{Wave Simulation}
\textbf{Data placement:} Wave simulation largely operates over arrays and employs \cedit{three types of operand interaction: elementwise, reduction, and neighboring mesh element.} 
Of these, while elementwise interaction can be tackled via data placement as was done for \textit{vector-sum}, reduction and neighbor interaction warrant more care. For the former, blocked data placement is employed (discussed below in the context of \textit{ss-gemm} primitive which also employs this). For the latter, array (grid) elements are placed such that, to the extent possible, neighboring faces reside in the same bank as depicted in Figure~\ref{fig:orch_workloads}b.

\textbf{Command orchestration:} Despite their PIM amenable properties, \textit{wavesim} primitives exhibit complex interaction patterns between operands which complicate orchestration. Considerable care is necessary to effectively utilize available registers while avoiding memory spills and lowering row activation overheads. While we hand schedule the computation in our analysis, 
\revise{existing compiler methods for register allocation~\cite{chow1990priority,briggs1994improvements} can be adapted for PIM-specific cost awareness.}

\begin{figure}[t]
    \centering
    \includegraphics[width=\linewidth]{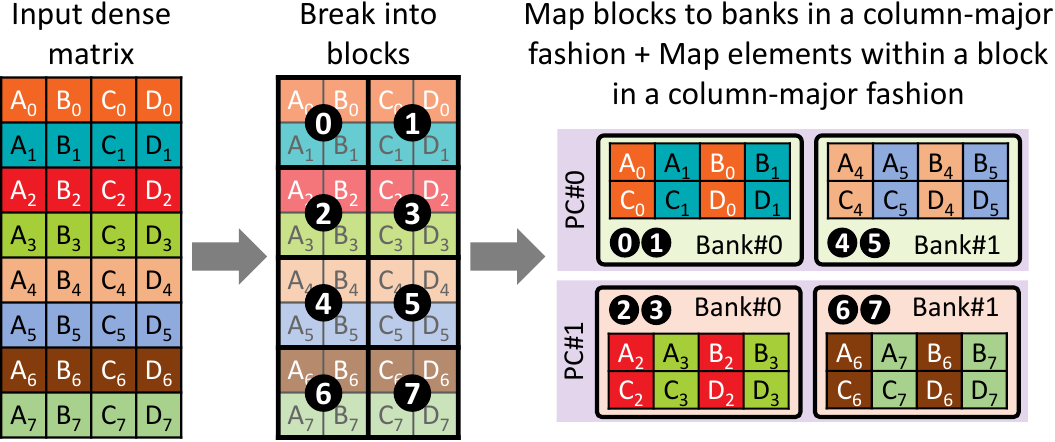}
    \caption{Workload \textit{ss-gemm} data mapping.}
    \label{fig:ssgemm_datamapping}
    \vspace{-\baselineskip}
\end{figure}

\putsssec{pim_base_offload_ssgemm}{Sparse Skinny GEMMs}
\textbf{Data placement:} To harness broadcast \textit{pim-commands} and concomitant performance, we place \cedit{the} input dense matrix in a blocked format as depicted in \figref{fig:ssgemm_datamapping}, which is tailored to both the dimensions of the matrix and the address interleaving of the system. This layout has multiple attractive properties: skinny matrix values can be broadcast as an immediate PIM operand on the data bus, and accumulation of partial products avoids inter-bank, intra-SIMD, and (to the extent possible) inter-row operations.

\textbf{Command orchestration:} Compute orchestration follows from data placement: an element of \cedit{the} skinny matrix is broadcast to banks and partial results are accumulated in \textit{pim-registers} before being written to memory. 

\putsssec{pim_base_offload_push}{Push-based Computation}
\textbf{Data placement:} Variation in graph connectivity precludes \cedit{the} use of broadcast commands and co-location of interacting neighbors (source and destination nodes in Figure~\ref{fig:orch_workloads}d). Instead, single-bank \textit{pim-commands} execute in-place destination updates, avoiding operand locality layout constraints.

\textbf{Command orchestration:} Compute orchestration for \textit{push-primitive} also follows from its data placement. The source node's value is read from memory, then updates to neighboring nodes are calculated and applied using single-bank \textit{pim-commands} (namely, a \textit{pim-ADD} command loads the current value and adds an operand supplied on the data bus, \cedit{placing} \cedit{the} result in a \textit{pim-register}, and a \textit{pim-store} command stores this result to memory).

\putssec{pim_base_perf}{Performance Analysis}
\begin{table}[t]
\centering
\caption{Parameters for performance model~\cite{HBM3-jedec}.}
\label{tab:model_params}
\resizebox{\columnwidth}{!}{%
\begin{tabular}{|l|l|}
\hline
\textbf{\#Banks per Channel/4-high Stack}      & 16 / 512                                 \\ \hline
\textbf{Bandwidth per Pin}               & 4.8 Gb/s                            \\ \hline
\textbf{GPU Mem. Bandwidth per Stack} & 614.4 GB/s                          \\ \hline
\textbf{Row Buffer Size}                 & 1024 B                              \\ \hline
\textbf{DRAM Parameters}                 & \makecell[l]{tRP=15ns, tCCDL=3.33ns, \\ tRAS=33ns} \\ \hline
\textbf{PIM Parameters} & \begin{tabular}[c]{@{}l@{}}\#PIM Units per Stack = 256\\ \#PIM Registers per ALU = 16\end{tabular} \\ \hline
\textbf{Peak HBM Bandwidth}                 & 614.4 GB/s                              \\ \hline
\end{tabular}%
}
\end{table}

We first discuss our performance model assumptions and follow that with analyzing the attained PIM acceleration for primitives under study. Finally, we conclude with discussing unique challenges that arise for computation offloads to PIM and also an opportunity which\cedit{,} if addressed and harnessed respectively, can unlock further PIM acceleration.

\putsssec{pim_base_perf_model}{Performance Models}
In our work, we use analytical models to evaluate performance. Our choice is guided by \cedit{the} following reasons. First, commercial PIM designs are still only available as functional prototypes. Second, we aim to study primitives considering realistic problem sizes, where PIM is likely to be beneficial. This renders GPU simulators difficult to use due to long simulation times. Finally, we assume a strong GPU baseline, even considering future software optimizations (see below). 

\textbf{GPU Performance Model:} For our GPU baseline, we assume the execution time is primarily a function of memory bandwidth (assumed to be 90\% of peak) and data accessed. Further, we assume perfect reuse with two exceptions: inter-timestep reuse is not modeled for \textit{wavesim} (we assume polynomial degree $p=2$, $729$ data points per element, and $65K$ elements per GPU, which is too large to fit in cache), and cache locality for \textit{push-primitive} is based on actual cache hit rates measured using rocprof~\cite{rocprof} with push-based workloads from graphBIG~\cite{nai2015graphbig} (specifically, hit rates of 44\%, 20\%, and 57\% are observed for roadnet-usa~\cite{davis2015suitesparse}, a synthetic power-law graph with 1M nodes and 10M edges, and a synthetic power-law graph with 10M nodes and 100M edges, respectively).

We believe this to be a fair assumption for memory-limited workloads which manifest low op\cedit{/}byte ratios as discussed in Section~\ref{sec:pim_amenability_workloads}. Further, we assume HBM3 memory~\cite{HBM3-jedec} in our analysis (Table~\ref{tab:model_params}). We do so to be both forward-looking and avail our baseline GPU with \cedit{the} best possible memory bandwidth. As such, this provisions a compelling baseline \cedit{against which to compare PIM acceleration benefits}. 

Further, for \textit{ss-gemm}, we assume an optimized GPU baseline which can exploit row-sparsity to both avoid loading the zero rows and computing on them. We estimate this sparsity by analyzing row-sparsity occurrence for computations in MLPerf DLRM-based recommendation model~\cite{naumov2019deep} using the Terabyte Click Logs testing dataset~\cite{criteo_logs}. 

\textbf{PIM Performance Model:} For PIM, we first deduce detailed \textit{pim-commands} (Section~\ref{sec:pim_base_offload}) and subsequently take into account DRAM timings (Table~\ref{tab:model_params}) and operations (row activation, etc.) to determine PIM execution time. Multi-bank \textit{pim-commands} are issued in-order at half the rate\footnote{As dictated by the $t_{CCDL}$ timing parameter for back-to-back requests to the same bank group, as opposed to the minimum possible time between reads/writes: $t_{CCDS}$.}  of regular reads/writes as is the case with the HBM-PIM design~\cite{samsungPIM}. Single-bank \textit{pim-commands} can be freely reordered and can be issued at the same rate as regular reads/writes. Further, \textit{push-primitive} updates are also assumed to occur atomically, which can be guaranteed by existing per-address ordering assumptions in the memory controller.

\begin{figure}[t]
    \centering
    \includegraphics[width=\linewidth]{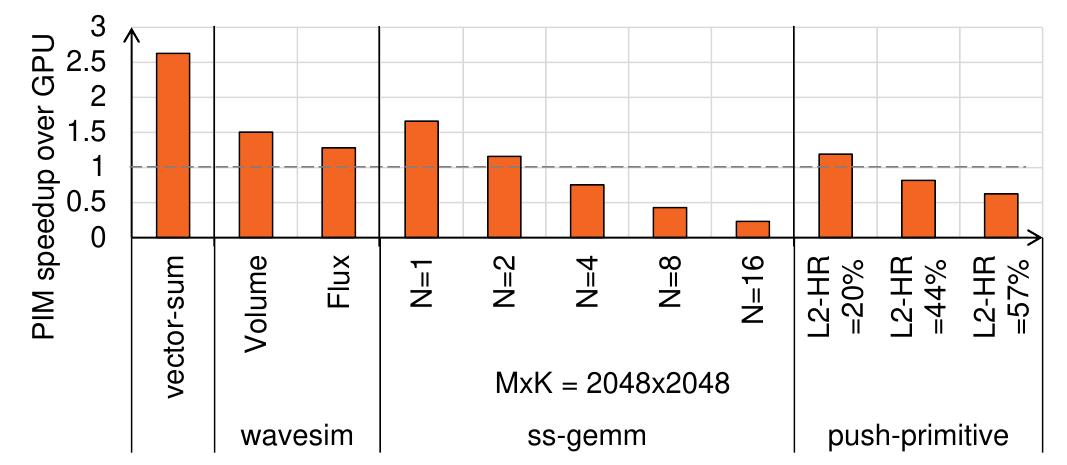}
    \caption{Commercial PIM speedup relative GPU. For \textit{ss-gemm}, N represents skinny matrix width. For \textit{push-primitive}, L2-HR represents the hit rate measured at L2 cache for each graph evaluated.} 
    \label{fig:baseline_pim_speedup}
    \vspace{-\baselineskip}
\end{figure}

\putsssec{pim_base_speedup}{PIM Speedup Analysis}
We depict PIM speedups for studied primitives in Figure~\ref{fig:baseline_pim_speedup}. For our PIM strawman design, the upper bound for performance is about 4x assuming the baseline GPU can utilize 100\% of peak memory bandwidth (optimistic). As depicted, while \textit{vector-sum} attains over 2.6x speedup, primitives under study in this work fare far worse, delivering between 0.23x-1.66x speedups vis-a-vis GPU. 

For \cedit{the} \textit{ss-gemm} primitive, except very skinny matrices ($N=2$),  PIM incurs increasing slowdown (between 25-77\%) as $N$ increases compared to the GPU. This is expected because, as data reuse improves, moving data to the GPU and exploiting the reuse on chip is more beneficial. Similar trends are observed for \textit{push-primitive} as cache hit rate improves. Overall, we observe that, even with considerable care to place data appropriately and orchestrate computation efficiently, commercial PIM designs do not attain broad acceleration. 

\putsssec{pim_base_challenges}{Challenges/Opportunity for PIM Acceleration}
We discuss in this section further analysis of our PIM speedups\cedit{,} focusing on unique challenges when computations are offloaded to PIM and also a unique opportunity. We also discuss how the identified challenges are more a function of commercial PIM designs and not specific to primitives under study. That is, addressing these challenges will likely enable broad acceleration with commercial PIM. 

\textbf{Challenge - Row-activations:} A key impediment to PIM acceleration for all primitives (including \textit{vector-sum}) is row activation overheads - that is, latency costs to open DRAM rows. In fact, this accounts  for 27\% and 50\% of total latency for \textit{wavesim-volume} and \textit{wavesim-flux} primitives. In the baseline GPU, bank-level parallelism, wherein row activation latency in one bank is overlapped with data access from another bank, can be employed. However, this is not possible for multi-bank \textit{pim-commands}. Further, while \textit{pim-registers} can be used to stage data from rows to lower row activations, this is difficult to realize for primitives with complex operand interaction\cedit{s} and high intermediate results which also consume registers (e.g., \textit{wavesim} primitives).

\textbf{Challenge - Cache reuse:} Primitives whose access patterns exhibit high cache locality are a poor fit for PIM, since on-chip reuse exploited by the GPU can outweigh the bandwidth benefit of PIM. In some cases, statically determining reuse potential and/or which accesses will exhibit cache reuse is difficult because locality and reuse potential is dependent on the input data (e.g., for \textit{ss-gemm} and \textit{push-primitive}). With a binary PIM-offload policy (all or nothing), while \cedit{the} GPU can harness \cedit{the} benefits of locality, baseline PIM does not\cedit{,} leading to poor performance. 

\textbf{Challenge - Registers/command bandwidth:} We also observe in this work that PIM architecture decisions, specifically the number of registers and command bandwidth available to single-bank \textit{pim-commands}, can limit PIM acceleration. While we discussed register space implications above, command bandwidth is a unique bottleneck for \textit{push-primitive} which relies on single-bank PIM commands. While \cedit{the} regular GPU access rate is limited by data bandwidth, and multi-bank PIM command rate is limited by ALU resources, single-bank PIM command rate, in the case of \textit{push-primitive}, is exclusively limited by command bandwidth especially since this primitive relies on commands which do not send operands on the data bus (\textit{pim-store}, Section~\ref{sec:pim_base_offload_push}).

\textbf{Opportunity - Sparsity:} For our \textit{ss-gemm} GPU implementation, we assume a GPU baseline that can exploit sparsity. If, similarly, PIM can exploit sparsity, perhaps even finer-grain sparsity than is possible at \cedit{the} GPU, \cedit{PIM acceleration can be improved further.} 
To this end, we observe that commercial PIM designs orchestrate computation using \textit{pim-commands} issued subject to fixed timing constraints \aedit{(}similar to how traditional memory operations are issued\aedit{)}. 
These commands are issued at fine granularity (a \textit{pim-command} accesses at most one 32B DRAM word per bank). This, as we show below, can be used to exploit fine-grain \aedit{data sparsity} with PIM.

%% file: 06-pim-optimizations.tex
\vspace{-2pt}
\putsec{pim-opt}{Optimized PIM}

\begin{figure*}[!t]
    \centering
    \includegraphics[width=\textwidth]{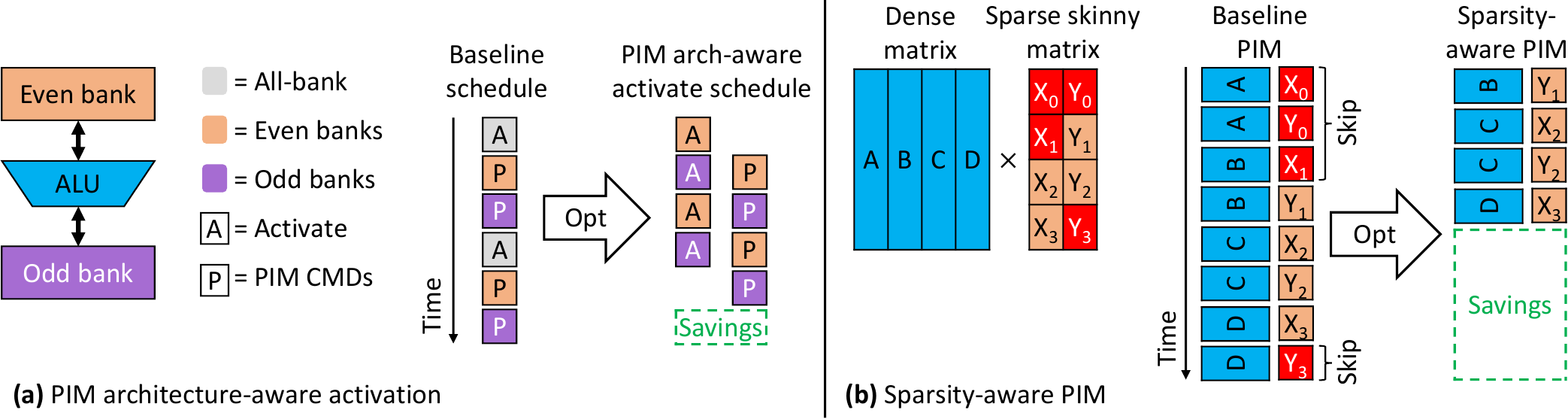}
    \caption{Optimizations for improving commercial PIM performance.}
    \label{fig:pim_optimizations}
    \vspace{-\baselineskip}
\end{figure*}

We discuss in this section some optimizations which can help improve acceleration availed by commercial PIM across varying domains and subsequently analyze the impact of these optimizations on PIM acceleration.


\putssec{pim_opt_list}{Targeted PIM Optimizations}

We discuss here both hardware augmentations and software optimizations that can help address challenges and harness opportunity discussed in Section~\ref{sec:pim_base_challenges}.

\putsssec{arch_aware_row_activation}{PIM Architecture-aware}
We first tackle row activation overhead evident in PIM executions. To this end, we design a PIM architecture-aware optimization, which we term \textit{architecture-aware row activation}. In \aedit{the} baseline PIM design, row activation for all banks is incurred on the critical path \cedit{and is} followed by compute commands\cedit{, each of which are issued to a subset (even or odd) of the available banks
(see \textit{Baseline schedule} in Figure~\ref{fig:pim_optimizations}a), 
as the PIM unit/ALU is shared by two DRAM banks.} 
Note that keeping PIM compute commands in program order is necessary to honor register dependencies. 

Instead of this schedule, we propose to first split all-bank row activations into separate even and odd bank activations. This allows for a decoupled parallel schedule depicted as \textit{PIM arch-aware schedule} in Figure~\ref{fig:pim_optimizations}a. 
\cedit{This schedule hides activation latency behind useful work by eagerly activating the next row in one subset of banks while compute commands are being performed in the opposite subset.} 
It also does not impact the order of compute commands or serialized activation latencies and dependencies within odd and even banks, which ensures functionally correct execution. Memory controllers can be augmented to generate/issue such architecture-aware row activations. Further, a compiler pass can also aid in \cedit{the} generation of architecture-aware row activations.

\vspace{-2pt}
\putsssec{sparsity_aware_pim}{Sparsity-aware}
A unique opportunity that commercial PIM designs avail (Section~\ref{sec:pim_base_challenges}) is thanks to \cedit{their} reliance on \textit{pim-commands} issued at fine-granularity subject to fixed DRAM timing constraints (similar to how traditional memory operations are issued). This design decision can be harnessed at \aedit{the} software level to perform additional checks at \cedit{the} processor to opportunistically skip issuing commands when certain conditions (e.g., sparsity) are met. Figure~\ref{fig:pim_optimizations}b illustrates how this works 
for \textit{ss-gemm}; before issuing a \textit{pim-command} to multiply a value of the sparse skinny matrix with a column in the dense matrix, the sparse matrix value is inspected; if this value is zero, then the command is skipped, 
\aedit{improving} performance. 

Our above \textit{sparsity-aware} PIM optimization has several notable benefits. First, this functionality enables PIM to harness dynamic sparsity (execution time sparsity) which is typically hard to exploit \cedit{in} GPUs~\cite{zhu2019sparse}. Further, our proposed functionality allows harnessing of sparsity without reliance on specialized sparsity formats which incur data-transformation and metadata overheads. Finally, \textit{sparsity-aware} PIM allows exploiting of {\textbf{finer}}-grain sparsity than possible for GPU.  The PIM design can \aedit{easily} exploit element-level sparsity whereas GPU harnesses row-level sparsity: skip loading rows with all zeros for the skinny matrix (note, dynamic element-level sparsity at \aedit{the} GPU will require creating specialized sparsity-aware data formats at runtime). 


\putsssec{cache_aware_pim}{Cache-aware}
A binary PIM offload decision hurts PIM performance (\textit{push-primitive}) by obviating input-dependent caching benefits for PIM executions (Section~\ref{sec:pim_base_challenges}). Instead, we believe that selective offloads to PIM \aedit{that consider} cache reuse is a more superior strategy. To this end, we make a case for \textit{cache-aware} PIM optimization, where existing fine-grain \textit{pim-instructions} are offloaded to PIM under the guidance of a locality predictor. Note that such techniques have been studied by prior works (\aedit{e.g.,} dynamic schemes~\cite{Ahn15}, offline schemes~\cite{heterographcache2018}) and can be augmented to work with commercial PIM designs. 

\vspace{-2pt}
\putsssec{limit_studies}{Limit Studies}
While so far we have discussed targeted optimizations which can broaden commercial PIM acceleration reach, we also believe that PIM architecture decisions, 
specifically the command bandwidth available to single-bank \textit{pim-commands} and the number of PIM registers, 
are important determinants of PIM acceleration. To that end, we also study in this work the effects of varying these on PIM performance. We believe that careful attention to these decisions will be necessary to avail broad PIM acceleration. 

\putssec{pim_opt_perf}{Performance Analysis}
We evaluate the implications of optimizations and techniques we discussed above on PIM performance below. Optimizations are employed largely in a \textit{targeted} manner, focused towards the primary bottlenecks of each primitive. 
\aedit{
Since \textit{wavesim} primitives exhibit register pressure that exacerbates activation overhead, we study \textit{architecture-aware} row activation and increased register resources for these primitives.
Since \textit{ss-gemm} exhibits dynamic sparsity, we study \textit{sparsity-aware} PIM for this primitive.
Since \textit{push-primitive} exhibits input-dependent cache locality and is limited by command bandwidth, we study \textit{cache-aware} PIM and increased command bandwidth resources for this primitive.
}
While these bottlenecks 
are specific to primitives under study, 
we do believe that these are fundamental challenges which will be experienced more widely as PIM is harnessed more widely. 
\aedit{Also note that these optimizations are complementary and can be employed in tandem in future inclusive PIM designs.}

\putsssec{pim_opt_wavesim}{Wave Simulation}
\begin{figure}[t]
    \centering
    \includegraphics[width=\linewidth]{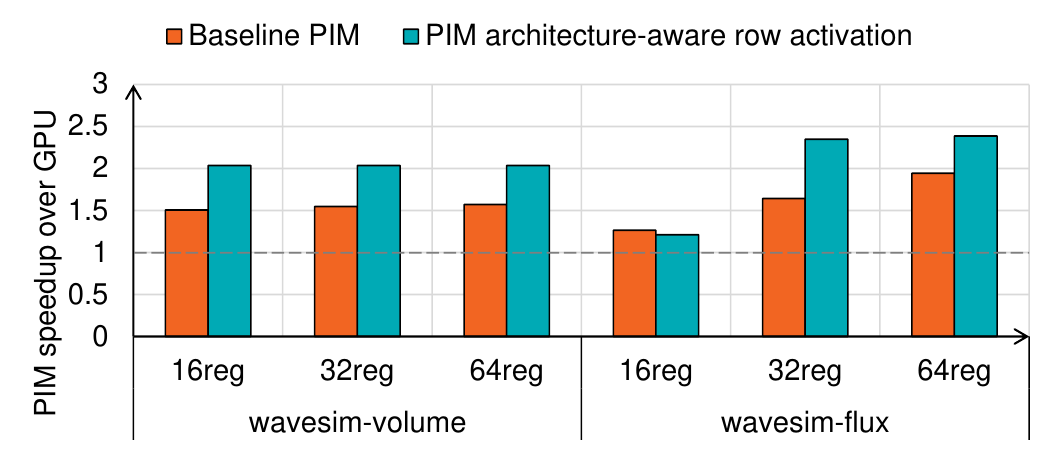}
    \caption{Optimized PIM speedup for \textit{wavesim} primitives.}
    \label{fig:wavesim_opt_pim_speedup}
    \vspace{-\baselineskip}
\end{figure}

Wavesim acceleration improvements with \textit{architecture-aware} row activation and increased register resources\footnote{Baseline PIM in line with commercial PIM designs assumes sixteen registers.} are depicted in Figure~\ref{fig:wavesim_opt_pim_speedup}. For \aedit{the} \textit{wavesim-volume} primitive, \textit{architecture-aware} optimization improves PIM speedup from 1.5x to 2.04x. Further, this optimization entirely eliminates row activation overheads for this primitive \aedit{such that more registers do not improve performance}.
This is in contrast to \aedit{the} \textit{wavesim-flux} primitive which exhibits higher register pressure and row activation overheads. At lower register counts (16), \textit{architecture-aware} activation does not improve performance because there are not enough commands per row activation to hide parallel activation latency or to amortize serial activation latency. However, \aedit{more resources reduce register pressure, enabling this optimization to better hide activation latency and achieve}
up to $2.63\times$ speedup over \cedit{the} GPU baseline.

\putsssec{pim_opt_ssgemm}{Sparse Skinny GEMMs}
\begin{figure}[t]
    \centering
    \includegraphics[width=\linewidth]{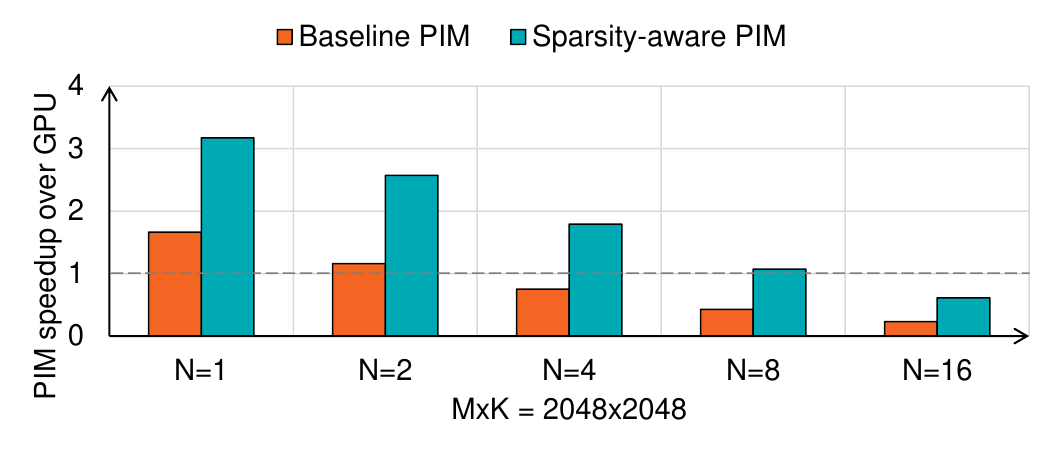}
    \caption{Optimized PIM speedup for \textit{ss-gemm}.}
    \label{fig:ssgemm_opt_pim_speedup}
    \vspace{-\baselineskip}
\end{figure}

For \textit{ss-gemm}, we focus on implications of our \textit{sparsity-aware} PIM optimization (depicted in Figure~\ref{fig:ssgemm_opt_pim_speedup}). We observe here that \textit{sparsity-aware} PIM significantly improves the PIM speedup (more than $3\times$) with expected tapering in benefits with increased reuse at \aedit{the} GPU (increasing $N$). Further, it also allows PIM to manifest acceleration in scenarios where baseline PIM manifested a slowdown (speedup of $1.07\times$ for $N=8$, while baseline PIM suffers from 57\% slowdown).

\putsssec{pim_opt_push}{Push-based Computation}
\begin{figure}[t]
    \centering
    \includegraphics[width=\linewidth]{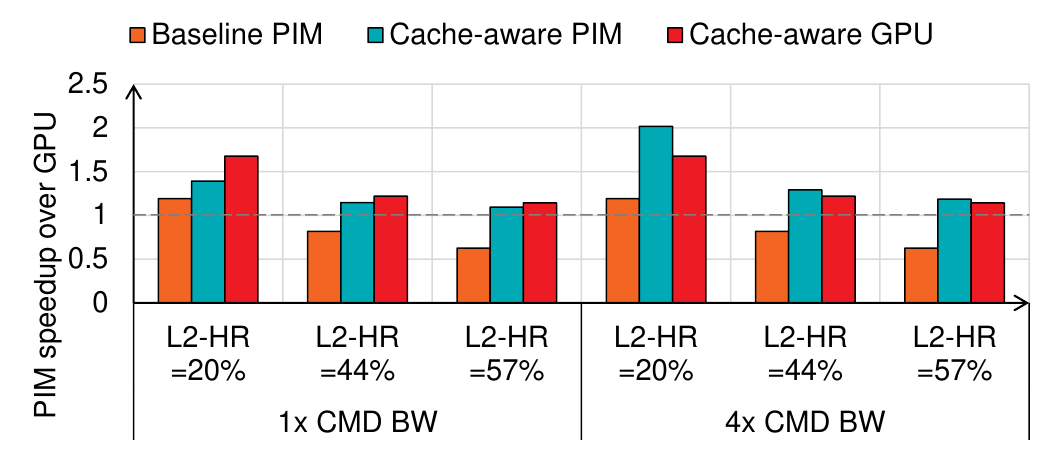}
    \caption{Optimized PIM speedup for \textit{push-primitive}.}
    \label{fig:graph_opt_pim_speedup}
    \vspace{-\baselineskip}
\end{figure}

As PIM acceleration for \textit{push-primitive} is \aedit{most} hindered 
from cache unawareness and lower command bandwidth for single-bank \textit{pim-commands}, we preferentially study the effects of both of these for this primitive and depict the resulting acceleration in Figure~\ref{fig:graph_opt_pim_speedup}. For this study, we model the effects of \cedit{a} locality-based predictor, using a cache model (16-way, 4MB, LRU replacement) which classifies updates to graph nodes in \textit{push-primitive} as either likely manifesting reuse (performed in cache) or not (performed in PIM). As before, we evaluate these workloads on three graph inputs with varying degrees of locality, each of which is labeled with its observed L2 cache hit rate. 
Overall, \aedit{\textit{cache-aware} PIM prevents performance degradation related to the cache reuse observed in the baseline PIM,} 
leading to an average speedup of $1.20\times$ (max $1.39\times$).

Further, we also model an optimized GPU baseline wherein the GPU can also leverage the locality predictor to reduce access granularity (i.e. use 32B rather than 64B accesses) for updates that do  not benefit from caching. We term this \textit{cache-aware GPU}, and it achieves up to $1.68\times$ speedup relative \aedit{to} \cedit{the} baseline GPU. Although \textit{cache-aware} PIM reduces data transferred across the memory interface relative \aedit{to} \textit{cache-aware} GPU, it does not reduce the command bandwidth demand, and the strictly modeled DRAM latency requirements actually lead to worse performance for \textit{cache-aware} PIM. 


Additional command bandwidth only benefits single-bank PIM commands that do not carry data (i.e., \textit{push-primitive} \textit{pim-store} commands).
With $4\times$ as much command bandwidth,\footnote{A 4x multiplier roughly corresponds to how much additional command bandwidth is possible if underutilized data bandwidth could carry command information.}
PIM performance improves further to exceed \textit{cache-aware} GPU performance for all inputs and provide up to $2.02\times$ speedup relative \aedit{to} \cedit{the} baseline GPU.

Overall, while the introduction of commercial PIM designs reveals new bottlenecks for many primitives, there clearly  is opportunity to address these bottlenecks to support a broader range of primitives.

%% file: 07-discussion.tex
\vspace{-2pt}
\putsec{discussion}{Discussion}

This section details additional PIM design considerations that we don't include in our evaluation, but which are nevertheless important for efficient PIM execution.

\ignore{
\textbf{Generality of findings:}
Although this work is focused on a specific PIM architecture and small set of primitives, the insights can apply more broadly. Row activation overhead can limit performance for any DRAM-based PIM architecture where strict command ordering is needed to preserve dependencies on near-memory intermediate storage (i.e., PIM registers), and this is exacerbated when workloads exhibit high register pressure and near-memory storage resources are scarce.

On-chip reuse is an important consideration for any PIM architecture, which only provides benefit for accesses that miss in the cache.
Sparsity-aware and locality-aware orchestration can be leveraged to enhance PIM benefits for workloads with input-dependent reuse, performing operations in PIM only when they are necessary and only when they cannot be performed more efficiently on-chip.

Command bandwidth can become a bottleneck for any PIM architecture that operates by issuing fine-grain commands to memory which are broadcast to multiple memory modules in parallel. In such systems, workloads that exhibit non-aligned parallelism can reduce memory traffic through the use of unicast PIM requests, but performance improvements are limited due to the command bandwidth bottleneck. PIM systems that can increase effective command bandwidth can deliver gains for any such workloads.
}

\textbf{Secondary PIM benefits:}
While we focus on performance benefits, PIM execution offers secondary benefits such as power efficiency and compute utilization. Power benefit arises from the reduction in data movement enabled by PIM. All studied primitives reduce how much data is transferred across the memory interface 
(by up to $71\%$ for \textit{wavesim-volume}, $98\%$ for \textit{wavesim-flux}, $87\%$ for \textit{ss-gemm}, and $46\%$ for \textit{push-primitive}),  and all but \textit{push-primitive} reduce how many memory requests are sent from the GPU to memory. In addition, PIM kernels do not use GPU compute resources and use relatively low memory bandwidth to issue PIM commands. PIM therefore also enables co-scheduling of complementary (e.g., compute-intensive) kernels on the GPU. 

\textbf{Data precision:} Since current commercial PIM designs are geared towards error-tolerant ML workloads, they do not support higher precision operations than FP16.
This can pose challenges for scientific and graph analytics domains, which often assume higher precision data types. 
Studies like this work can potentially motivate commercial PIM support for higher precision arithmetic, which is one path to resolving this disconnect. 
However, there have also been multiple recent efforts to enable the use of low-precision arithmetic for HPC and graph workloads~\cite{numericalmethods2020,harnessingTCforIterative2018,meforhpc2021,modularprecision2019}. Although these efforts generally aim to reduce the memory footprint or leverage the low-precision matrix engines that exist in ML-optimized GPUs, many of the associated insights can apply to (and are further motivated by) commercial PIM designs as well.

\textbf{Application-level considerations:}
Although our \textit{PIM-amenability-test} (Section~\ref{sec:pim_amenability_test}) is designed to be generally applicable, our evaluation of individual PIM primitives does not consider overheads that arise when 
PIM and non-PIM kernels interact.
Commercial PIM systems assume that the target data is present in memory, which can require cache flushes or the use of uncached memory buffers to ensure PIM consistency when communicating between PIM and non-PIM kernels. The associated overheads of such requirements (flush traffic and latency, copies to uncached buffers, and/or reuse prevented by these actions) should be incorporated in the cache reuse and memory residency test when evaluating a primitive in the context of a larger application.


Similarly, kernel fusion is a common optimization wherein multiple computations are fused to avoid round trips to memory. Note that such fusion can be employed for PIM offloads too. If offloading to PIM prevents kernel fusion for GPU, this cost should be factored in assessing PIM-amenability. 


\textbf{Address hashing:} 
Many systems use address hashing to improve memory efficiency by effectively shuffling how addresses are mapped to physical memory units.
However, this can conflict with PIM's need for control over data placement to ensure operand locality and, 
depending on the hashing function, may even preclude PIM execution. 
Possible solutions include disabling hashing when PIM is enabled, exposing hashing to software (e.g., page coloring to identify the same hashing group), or limiting address hashing to bits that are irrelevant to PIM mapping 
(e.g., only hash bank bits which are shared by one PIM unit or row bits only).

\textbf{Memory reliability, security:} 
Error detection and correction (EDC) are important features that help ensure the integrity of DRAM access, but incur overhead to verify DRAM loads and set the proper metadata bits for DRAM stores.  Similarly, memory encryption/decryption may be needed for privacy-sensitive systems, requiring similar encryption/decryption on the path to and from memory.
If EDC or encryption are to be supported with PIM, these functions must be replicated near-bank, requiring additional area overhead. 
Though they can be pipelined to avoid impacting bandwidth, they also add latency that can appear on the critical path of PIM operations that access memory. 
However, in many cases this latency can be hidden by interleaving PIM operations to independent parallel data, in the same way that registers are used to exploit row locality.


%% file: 08-related.tex
\putsec{related}{Related Work}

Many recent efforts study compute orchestration on different PIM architectures.
In this section we summarize some of the most relevant past studies not previously discussed.

In contrast to the near-bank coupling of HBM-PIM and GDDR-PIM, many recent PIM architectures are based on compute units implemented on a "base" logic die 3D-stacked under a set of DRAM dies~\cite{boroumand2019conda,boroumand2016lazypim,choe2019concurrent,zhang2014top,nair2015active,ahn2015pim,boroumand2018google}. However, as HBM already provisions high bandwidth to an external processor via 2.5D-stacking, these 
approaches do not provide sufficient bandwidth amplification in the context of HBM.

Some prior approaches have proposed application-specific capabilities on DDR DIMMs 
~\cite{kwon2019tensordimm,ke2020recnmp}. However, our focus is on extending in-memory capabilities to a broader set of workloads. Some prior efforts on broader PIM acceleration incorporate full-blown compute cores in DDR DRAM~\cite{devaux2019true}. However, this approach has a higher area cost in DRAM compared to the approaches we consider here due to the presence of instruction fetch and sequencing hardware. 

Despite \cedit{the} above-mentioned differences, several of the prior PIM approaches can also be considered complementary to this work as they do not impact DRAM bank architecture and can be deployed in tandem with  HBM-PIM-like techniques.

Some PIM architectures integrate compute and memory more tightly than HBM-PIM and GDDR-PIM.
These architectures attain extreme parallelism by executing bulk bitwise functions directly on bitline outputs~\cite{li2017drisa,seshadri2017ambit} or by leveraging the physical properties of non-volatile memory (NVM) to perform analog operations~\cite{mittal2018survey,li2022survey,xi2020memory}.
However, these systems are more limited in the types of primitives they can accelerate. While ReRAM efficiently implements dot products for weight-stationary inference and bitwise operations can be chained to implement general arithmetic, 
\aedit{the accuracy, precision, and programmability of these systems are limited, precluding their use for many compute domains.} 
When such architectures are leveraged for irregular workloads (e.g., GraphR~\cite{song2018graphr}) the focus is energy and area efficiency rather than improved data bandwidth. 

Past work has also studied the role of various forms of PIM for accelerating wave simulation~\cite{wavePIM2021}, graph analytics~\cite{graphpimAhn2015,graphp2018,graphq2019,graphpimNai2017}, sparse ML~\cite{giannoula2022sparsep}, and many other primitives~\cite{gomez2021benchmarking}.
However, 
each of these targets a PIM architecture that differs in significant ways (they use domain-specific or loosely-coupled architectures) from the commercial PIM designs studied in this work. 

%% file: 09-conclusion.tex
\vspace{-2pt}
\putsec{concl}{Conclusion}

To the best of our knowledge, this is the first work to evaluate emerging commercial PIM designs across primitives from a broad set of domains. To this end, we first deduce a PIM-amenability-test, which can be used to assess \aedit{PIM's potential for accelerating primitives, 
and also to guide efficient orchestration and data-placement.}
\aedit{Using this test}, we observe that commercial PIM designs, which today are rightly geared toward (a narrow set of) ML primitives, 
do not accelerate \aedit{the} primitives under study even though these primitives \revise{exhibit PIM-amenable properties}. 
To address this, we identify bottlenecks unique to these PIM designs along with hardware/software optimizations that overcome these bottlenecks, improving average PIM speedups from 1.12x to 2.49x relative \aedit{to} a GPU baseline. Overall, our work shows that, while emerging commercial PIM designs hold promise, for broad acceleration, a more inclusive PIM design will be necessary.